\documentclass[iop]{emulateapj}
\usepackage{amssymb,amsmath}
\usepackage{subfigure}

\begin{document}

\newcommand{\halpha}{H\ensuremath{\alpha}}
\newcommand{\hbeta}{H\ensuremath{\beta}}
\def\msun{{\rm\,M_\odot}}

\title{Investigating {\halpha}, UV, and IR star-formation rate diagnostics for a large sample of $\lowercase{z}\sim 2$ galaxies}
\author{\sc Irene Shivaei\altaffilmark{1,2}, Naveen A. Reddy\altaffilmark{1,3}, Charles C. Steidel\altaffilmark{4}, Alice E. Shapley\altaffilmark{5}}
 
\altaffiltext{1}{Department of Physics and Astronomy, University of California, Riverside, 900 University Avenue, Riverside, CA 92521, USA}
\altaffiltext{2}{NSF Graduate Research Fellow}
\altaffiltext{3}{Alfred P. Sloan Research Fellow}
\altaffiltext{4}{Cahill Center for Astronomy and Astrophysics, California Institute of Technology, 1216 E. California Blvd., MS 249-17, Pasadena, CA 91125, USA}
\altaffiltext{5}{Department of Physics \& Astronomy, University of California, Los Angeles, 430 Portola Plaza, Los Angeles, CA 90095, USA }

\slugcomment{DRAFT: \today}
  
\begin{abstract}
We use a sample of 262 spectroscopically confirmed star-forming
galaxies at redshifts $2.08\leq z\leq 2.51$ to compare {\halpha}, UV,
and IR star-formation-rate diagnostics and to investigate the dust
properties of the galaxies.  At these redshifts, the {\halpha}
line shifts to the $K_{\rm s}$-band. By comparing $K_{\rm s}$-band
photometry to underlying stellar population model fits to other
UV, optical, and near-infrared data, we infer the {\halpha} flux for
each galaxy.  We obtain the best agreement between {\halpha}- and UV-based
SFRs if we assume that the ionized gas and stellar continuum are reddened
by the same value and that the Calzetti attenuation curve is applied to both.
Aided with MIPS 24\,$\mu$m data, we find that an attenuation curve
steeper than the Calzetti curve is needed to reproduce the observed
IR/UV ratios of galaxies younger than 100\,Myr.  Furthermore, using
the bolometric star-formation rate inferred from the UV and mid-IR
data (SFR$_{\text{IR}}$+SFR$_{\text{UV}}$), we calculated the
conversion between the {\halpha} luminosity and SFR to be $(7.5\pm1.3)
\times 10^{-42}$ for a Salpeter IMF, which is consistent with the
\citet{kennicutt98} conversion. The derived conversion factor is
independent of any assumption of the dust correction and is robust to
stellar population model uncertainties.

\end{abstract}
\maketitle

\section{Introduction}
\label{sec:intro}

One of the most important diagnostics in understanding the evolution
of galaxies is the star-formation rate (SFR). The evolution of the SFR
of galaxies can give clues as to how galaxies were enriched with heavy
elements, how they build up their stellar mass through cosmic time,
and helps us to understand the bolometric output of galaxies.  At
redshift $z\sim 2$, when the universe was just $\sim$ 3\,Gyr old,
star-formation activity in the universe was at its peak and galaxies
were in the process of assembling most of their stellar mass
\citep[see][]{reddy09,bouwens10,shapley11,madau14}. Studying this
critical epoch is essential to gaining a better understanding of the
evolution of the progenitors of the local galaxy population.

The ultra-violet (UV) continuum (1500 to 2800\,\AA\,) intensity of a
galaxy is one of the most commonly used diagnostics for the SFR as it
is observable over a wide range of redshifts and intrinsic
luminosities. It is sensitive to massive stars ($M_{\ast}\ga
5\,\msun$), making it a direct tracer of current SFR.  By
extrapolating the formation rate of massive stars to lower masses, for
an assumed form of the initial mass function (IMF), one can estimate
the total SFR \citep{madau98}. Another widely used diagnostic for
measuring the SFR is nebular emission, with {\halpha} being the most
common because of its higher intensity compared to the other hydrogen
recombination lines such as {\hbeta}, Pa$\alpha$, Pa$\beta$, etc., and
it is easier to interpret than the Ly$\alpha$ line.  {\halpha} is an
``instantaneous'' tracer of SFR because it is sensitive only to the
most massive stars ($M_{\ast}\ga 10\,\msun$).  However, it becomes
more challenging to observe {\halpha} from the ground at $z \ga 1$
because the line is redshifted to the near-IR where the terrestrial
background is much higher than at optical wavelengths.

The main disadvantage of using UV/optical luminosities as tracers of
the SFR is their sensitivity to dust attenuation. The dust absorption
cross-section is larger for shorter wavelengths and choosing the
appropriate attenuation curve to correct the observed luminosities
plays an important role in determining intrinsic physical
quantities. Aside from the assumed attenuation curve, the geometry of
dust with respect to the stars can lead to different color excesses,
$E(B-V)$, between the ionized gas and the stellar continuum.
$E(B-V)$ is
the color excess measured between the {\em B} and {\em V} bands,
$E(B-V) \equiv A_{B} - A_{V}$, where $A_{\lambda}$ is the total
extinction at wavelength $\lambda$ in magnitudes.  
In particular, the nebular recombination lines arise from the HII
regions around the most massive O and early-type B stars (with masses
of M$_{\ast}\ga 10\,\msun$ and main sequence lifetimes of $\la
10$\,Myr). On the other hand, for a Salpeter IMF, solar metallicity,
and a constant or rising star-formation history, the UV continuum in
starburst galaxies originates from stars over a broader range of mass
that includes later-type B stars with lifetimes $\la 100$\,Myr
\citep{kennicutt98,madau14}. These older non-ionizing stars have more
time to migrate to regions of lower dust density in the galaxy, while
H-ionizing stars with shorter lifetimes do not have enough time to
escape from their dusty birthplace or let the parent molecular clouds to dissipate.
As a result, the nebular lines can be subject to a higher degree of reddening than the UV
continuum.

\citet{calzetti94} found that the nebular emission is more attenuated
than the stellar continuum at the same wavelengths for a sample of
local UV-bright galaxies. Subsequently, \citet{calzetti00} studied a
similar sample of local galaxies and argued that the {\em
  color-excess} is 2.27 times larger for the nebular emission lines
than for the stellar continuum. This relationship was derived under the assumption that the Calzetti curve is applied to the stellar continuum and a Galactic extinction curve is applied to the nebular emission lines. In a separate study,
  \citet{fernandes05} found that for an SDSS sample of $\sim 50000$
  galaxies, $A_{V,nebular}\sim 1.8 ~A_{V,stellar}$, both assuming a
  \citet{cardelli89} extinction curve.

There have been several studies of relative attenuation of the stellar
continuum and nebular regions in high-redshift galaxies. 
In most of the high-redshift studies, the Calzetti curve is assumed for both the nebular and stellar regions. Frequently, 
the \citet{calzetti00} relation between the
stellar and nebular $E(B-V)$ is used to correct the nebular lines, while this relation was originally derived based on using a Galactic extinction curve for the nebular lines. At present it is unclear what type of attenuation curve should be applied to the nebular regions. These issues are discussed in detail in \citet{steidel14}.

At $z\sim 2$, \citet{forster09} selected a sample from various imaging
surveys in the optical, near-IR, mid-IR, and sub-millimeter regime,
with SED-derived SFRs of $\sim$ 10 - 1000 $\msun\,\text{yr}^{-1}$ and
stellar masses $\gtrsim 10^{10}\,\msun$, assuming a Chabrier IMF.  In
their study, the predicted {\halpha} luminosities from the best-fit
SED models for 62 galaxies were compared with the observed line
luminosities measured from integral field spectroscopy.  The latter
were then corrected for reddening assuming the $A_V$ returned from
stellar population modeling of their galaxies.  
\citet{forster09} found that assuming a factor of two redder
  color excess towards the HII regions relative to the stellar
  continuum yielded the best agreement between the predicted and
observed {\halpha} luminosities. Similarly, \citet{wuyts11}
investigated 25 star forming galaxies at $z\sim 2$ and compared the
{\halpha} SFRs, dust-corrected by the Calzetti attenuation curve, with
the SED modeled SFRs. They showed that {\halpha} SFRs are consistent
with the SED SFRs, provided that $E(B-V)_{neb} =
2.27 E(B-V)_{stellar}$ when assuming the Calzetti curve for both.

\citet{muzzin10} presented two galaxies at redshifts $z = 2.122$ and
2.024, and added ten other galaxies at similar redshifts from
\citet{vandokkum05}, \citet{kriek07}, and \citet{kriek09b} to compare
A$_{V,nebular}$ derived from the Balmer decrement
($\frac{F({\halpha})}{F({\hbeta})}$) with A$_{V,stellar}$ derived from
the SED models, both assuming the Calzetti curve. The only two
galaxies with detected H$\beta$ line measurements showed evidence for
higher nebular attenuation, but about half of the non-detections were
consistent with equal attenuation for the nebular and stellar regions.
In \citet{yoshikawa10}, {\halpha} attenuation was investigated through
several comparisons for a sample of optically-selected star-forming
galaxies at $z\sim 2$. Aided with near-infrared spectroscopic observations,
\citet{yoshikawa10} used the Balmer decrement along with the Calzetti curve 
to correct the {\halpha} SFR.
The dust-corrected SFR was then compared with the {\halpha} SFR
corrected by the SED color-excess and also with the SFRs inferred from
UV, IR, radio, and X-ray. The comparisons yield contradictory results.
Their analysis favored a higher nebular color-excess for galaxies with
larger intrinsic SFRs.

In separate studies, \citet{erb06c} and \citet{reddy10} used a large
sample of UV-selected galaxies at $z\sim 2$ and corrected both the
{\halpha} and UV SFRs with the color-excess derived from the best-fit
SED models and applied the Calzetti attenuation curve. They showed
that using different color excesses for the continuum and lines
generally resulted in {\halpha}-inferred SFRs that over-predicted
those calculated from the dust-corrected UV emission and direct IR
measurements.

The apparently conflicting results mentioned above may be reconciled
if the relation between the nebular and stellar color-excesses depends
on the physical properties of galaxies, such as their SFRs
\citep{yoshikawa10} or specific SFRs \citep{wild11}.
For example, 
\citet{wild11} showed that the ratio of the line optical depth to the
continuum optical depth decreases with increasing sSFR and at
sSFR$\sim 10^{-9}\,\text{yr}^{-1}$ the line-to-continuum optical
depth ratio reaches the \citet{calzetti00} ratio 
assuming the \citet{wild11} attenuation curve.
\citet{price14} investigated the
attenuation of the nebular regions at $z\sim 1.5$ using the Balmer
decrement from stacked {\em HST} grism spectra assuming the Calzetti
curve, and found no strong trend of
${\text{A}_{\text{v,neb}}}/{\text{A}_{\text{v,stellar}}}$ with SFR or
sSFR. 
In \citet{price14} study, at sSFR$\sim 10^{-8.5}\,\text{yr}^{-1}$
the ratio is consistent with 1.

The primary goal of this paper is to understand the relationship
between the UV and {\halpha} emission in high-redshift star-forming
galaxies, with a large sample that is immune to uncertainties in slit
loss corrections that affects the {\halpha} flux estimation \citep[c.f.][]{erb06c,yoshikawa10}, the small
sample sizes inherent in previous spectroscopic studies of {\halpha}
\citep{kriek07,muzzin10}, and not subject to the bias of selecting
high equivalent width objects from narrow-band selected samples
\citep{garn10}.  We consider in our
analysis only spectroscopically-confirmed galaxies, enabling us to
estimate the {\halpha} flux from broad-band photometric excess
techniques without the additional uncertainties that
plague photometric redshifts \citep{wuyts11}.

The impact of nebular lines on the broad-band photometry was known and studied for many years \citep[e.g.,][]{guiderdoni87,fioc&rocca97,zackrisson08,debarros14}.
Using the photometric excesses to determine the line strengths was 
pioneered by \citet{shim11}, where they showed that the excess in
{\em Spitzer}/IRAC 3.6\,$\mu$m relative to the SED model continuum is
due to the redshifted {\halpha} emission line for their sample of
galaxies at $3.8 < z < 5.0$. \citet{stark13} also investigated a
sample of galaxies at the same redshift range of \citet{shim11} and
inferred the {\halpha} emission line strengths by comparing the
observed flux in {\em Spitzer}/IRAC 3.6\,$\mu$m band and the continuum
flux as expected from the SED model. Following that,
  \citet{schenker13} verified the photometric excess technique by applying it to
  a small sample of 9 galaxies at $3.0 < z < 3.8$; for 8 galaxies the
  [O {\scriptsize III}] line fluxes inferred by the same technique as
  \citet{stark13} were within a factor of 2.5 of the spectroscopically
  measured [O {\scriptsize III}] fluxes.

An additional advantage of our study is that we include IR data 
to independently assess the dust-obscured SFR.
Comparing {\halpha}, UV, and IR-inferred SFRs in a statistical sense
allows us to understand how to correct extinction-sensitive measures
of SFR for the effects of dust.

The outline of this paper is as follows. In \S\ref{sec:sample} we
discuss the properties of our sample, the assumptions
that have been made to model the stellar populations using the rest-frame
UV to near-IR photometry,  
the {\it Spitzer}/MIPS photometry and stacking method. 
A detailed description of how we estimated the {\halpha} line flux
is provided in \S\ref{sec:sec3}. 
The analysis of the
MIPS 24\,$\mu$m data
and IR luminosities
is described in \S\ref{sec:irlum}.  In \S\ref{sec:sec5} we compare the two different
tracers of SFR - {\halpha} and UV - and discuss the dust correction
recipe most consistent with our measured values.
\S\ref{sec:sec6} focuses on 
combining SFR diagnostics (e.g., {\halpha} and UV with IR) to deduce
bolometric SFRs.  The results are summarized in \S\ref{sec:sum}.
Throughout this paper, a \citet{salpeter55}
IMF is assumed and a cosmology with H$_0$ = 70 km s$^{-1}$
Mpc$^{-1}$, $\Omega_{\Lambda}$ = 0.7, and $\Omega_m$ = 0.3 is
adopted. All magnitudes are given in the AB system
\citep{oke83}.

\section{Sample}  
\label{sec:sample}
\subsection{Sample Selection and Optical Photometry} 
\label{sec:sample-a}

The galaxies used in this study are drawn from a parent sample that is
part of an imaging and spectroscopic survey of UV-selected galaxies at
z $\sim$ 2-3 \citep{steidel04,reddy12b}.  The galaxies were selected
based on the BX, BM, and Lyman break galaxy rest-UV color criteria
\citep{steidel03,steidel04,adelberger04}, where $U_nGR$ optical data
were obtained with the Palomar Large Format Camera (LFC) or Keck Low
Resolution Imaging Spectrograph
\citep[LRIS;][]{steidel03,steidel04}. Rest-UV spectroscopic follow-up
with Keck/LRIS was conducted for galaxies brighter than $R = 25.5$
\citep{steidel03,steidel04}.  Near-IR $J$ and $K_s$ imaging was
obtained using the Palomar/WIRC and Magellan/PANIC instruments
\citep{shapley05,reddy12b}. {\em H}-band (F160w) data were obtained
with the {\em Hubble Space Telescope} WFC3 camera \citep{law12,
  reddy12b}. All galaxies in the sample have coverage in at least one
of the {\it Spitzer}/IRAC four channels (3.6, 4.5, 5.8, and 8.0~$\mu$m;
\citealt{reddy06a,reddy12b}).  The objects are located in the
GOODS-North field and 11 additional fields that were primarily
selected to have one or more relatively bright background QSOs for
studying the interface between the intergalactic medium (IGM) and
galaxies at z $\sim$ 2-3 \citep{steidel04,steidel10}.

Out of the final sample of 2283 objects with spectroscopically
confirmed redshifts, a subset of galaxies is selected based on the
following criteria: (1) the galaxy must be covered by the {\em K}-band
imaging, (2) it must have a redshift in the range 2.08 $\leq$ z $\leq$
2.51 so that the {\halpha} line falls into the {\em$K_{\rm s}$} band,
and (3) it must be detected in at least two of the IRAC channels or one of the IRAC
channels and either J or F160w bands. The third condition ensures a more
robust estimate of the stellar mass and the continuum level at
6564\,\AA\,. Furthermore, AGNs (making up $\approx 9\%$ of the parent
sample) were identified by either strong UV emission lines (e.g.,
Ly$\alpha$, CIV) or by a power law SED through the IRAC bands.  These
AGNs are removed from our sample. Eventually, 262 galaxies remain
that satisfy the aforementioned criteria.

\subsection{ Stellar Population Modeling}
\label{sec:sedfit}

For each galaxy in our sample, the best-fit stellar population model
is found by using the rest-frame UV through near-IR broad-band
photometry. As mentioned above, all the galaxies in our sample have
spectroscopically-confirmed redshifts, thus removing a key degeneracy
in the modeling of the stellar populations. In addition, for a better
estimation of stellar mass and age, all galaxies in our sample have at
least two detections long-ward of the 4000\,\AA\, break - excluding
{\em$K_{\rm s}$}-band.

C. Charlot \& G. Bruzual (2007) models with a \citet{salpeter55} IMF
and solar metallicities are used for the fitting. For each individual
galaxy, different star-formation histories are assumed, including
constant, exponentially declining, and exponentially rising with
characteristic timescales of $\tau$ = 10, 20, 50, 100, 200, 500, 1000,
2000, 5000\,Myr for exponentially declining and $\tau$ = 100, 200,
500, 1000, 2000, 5000\,Myr for exponentially rising histories. Ages
are allowed to vary between 50\,Myr and the age of the universe at the
redshift of each galaxy. The $>$ 50\,Myr limit corresponds to the
typical dynamical timescale of star-forming galaxies at $z\sim 2$ as
inferred from velocity dispersion and size measurements of these
galaxies \citep{reddy12b}. For interstellar dust obscuration, the
\citet{calzetti00} attenuation curve is used, allowing $E(B-V)$ to
vary between 0.0 and 0.6. The $\chi^2$ values have been determined for
each set of observed broad-band and model magnitudes. The best-fit
model is determined through $\chi^2$ minimization. There is generally
no significant difference between the best-fit $\chi^2$ values of the
six different population models (constant, exponentially rising, and
exponentially declining star-formation histories, for each considering
all ages and ages greater than 50\,Myr). As previous studies have
shown, the assumption of declining star-formation histories at these
redshifts results in systematically lower SED-inferred SFRs compared
to the observed IR+UV SFRs \citep{wuyts11,reddy12b}. Furthermore,
\citet{reddy12b} showed that assuming a constant star-formation
history for $z\sim 2$ galaxies predicts specific SFRs (SFR/M$_{\ast}$)
at higher redshifts ($z\sim 2.6$) that are substantially larger than
the observed values.  Given these, we adopt the models that assume
exponentially rising star-formation histories with ages greater than
50\,Myr.  

 The SED models used to fit the observed magnitudes do not include
  nebular emission lines. For example, the {\halpha} line can significantly affect the photometry and we use its
  contribution to the {\em K}-band to estimate the {\halpha} line
  flux. The [O {\scriptsize III}] emission line is the other strong line
  which falls into the F160w filter given the redshift range of our
  galaxies. Only 20\% of the galaxies have F160w observations, for which we did not correct the broadband
  photometry for the contamination. The SED-inferred SFR of these
  galaxies is consistent with the SFR(UV) estimates within the
  uncertainties.  At this redshift, J band is contaminated by the [O
    {\scriptsize II}] emission line, but this line is generally weaker
  than the {\halpha} line and its effect on the SED inferred SFRs is
  negligible compared to the uncertainties.\footnote{ For galaxies with similar SFRs   
  at $z\sim 2$, the
    typical [O {\scriptsize II}] line flux is $\sim 6\times
    10^{-17}$~erg~s$^{-1}$~cm$^{-2}$ \citep{kriek14}.
    The ratio of the [O {\scriptsize II}] flux to the typical J-band flux errors in
    our observations is only $\sim 0.07$.}

\subsection{MIPS data}  
\label{sec:sample-b}

To further investigate the bolometric properties of our sample, we
use {\it Spitzer}/MIPS 24\,$\mu$m wherever available.  Out of
12 fields, GOODS-North \citep{dickinsongoods03}, and four other fields
(Q1549, Q1623, Q1700, and Q2343; Reddy \& Steidel 2009) have MIPS
24\,$\mu$m coverage to a typical 3\,$\sigma$ depth of 10-15\,$\mu$Jy.

Photometry on 24\,$\micron$ images is performed by using point-spread
function (PSF) fitting with priors determined by the locations of the
objects in the higher resolution IRAC images (IRAC data exist in all
fields). A 40$\times$40 pixel region centered on each target is extracted 
with pixel size of $1\farcs 2$. PSFs are then
fitted simultaneously to all known sources in the sub-image and one
random background position. This procedure is repeated many times to
obtain sufficient statistics for proper background estimation based on
the random background flux measurements. The other source of
uncertainty is Poisson noise, which for objects in our sample is
negligible compared to the background dispersion. We remove objects
whose photometry may be compromised due to blending with nearby
sources, using the criteria specified in \citet{reddy10}. This results
in 115 galaxies with secure PSF fits.

Out of 115 objects with MIPS data, 47 are detected with S/N~$\le
3$. Undetected objects, those with S/N~$<3$, are considered using
either survival analyses or through stacking of the 24\,$\micron$
data.

\subsubsection{Stacking Method}

Throughout the paper, we employ a stacking method to determine the
median 24\,$\mu$m fluxes of objects that are individually detected and
undetected, following the procedures described in
\citet{reddy10,reddy12a}. We performed aperture photometry on the
stacked images, and applied an aperture correction based on the
$24$\,$\mu$m PSF. The average background level and noise were
determined by placing many apertures of the same size used for the
stacked signal at random positions in the stacked image and measuring
the average flux level and dispersion in flux of these ``background''
apertures.  Furthermore, we used bootstrap resampling simulations to
estimate the dispersion in the fluxes of objects contributing to each
stack.  This was accomplished by creating 100 samples of random images
in each bin and measuring the standard deviation of the median stacked
fluxes. 
The intrinsic dispersion in the stacked flux is larger than the
background error by a factor of $\sim 2$. 

\begin{figure}[tbp]
\subfigure{\includegraphics[trim=0cm 2.8cm 2.4cm 3.3cm,clip=true,width=.45\textwidth]{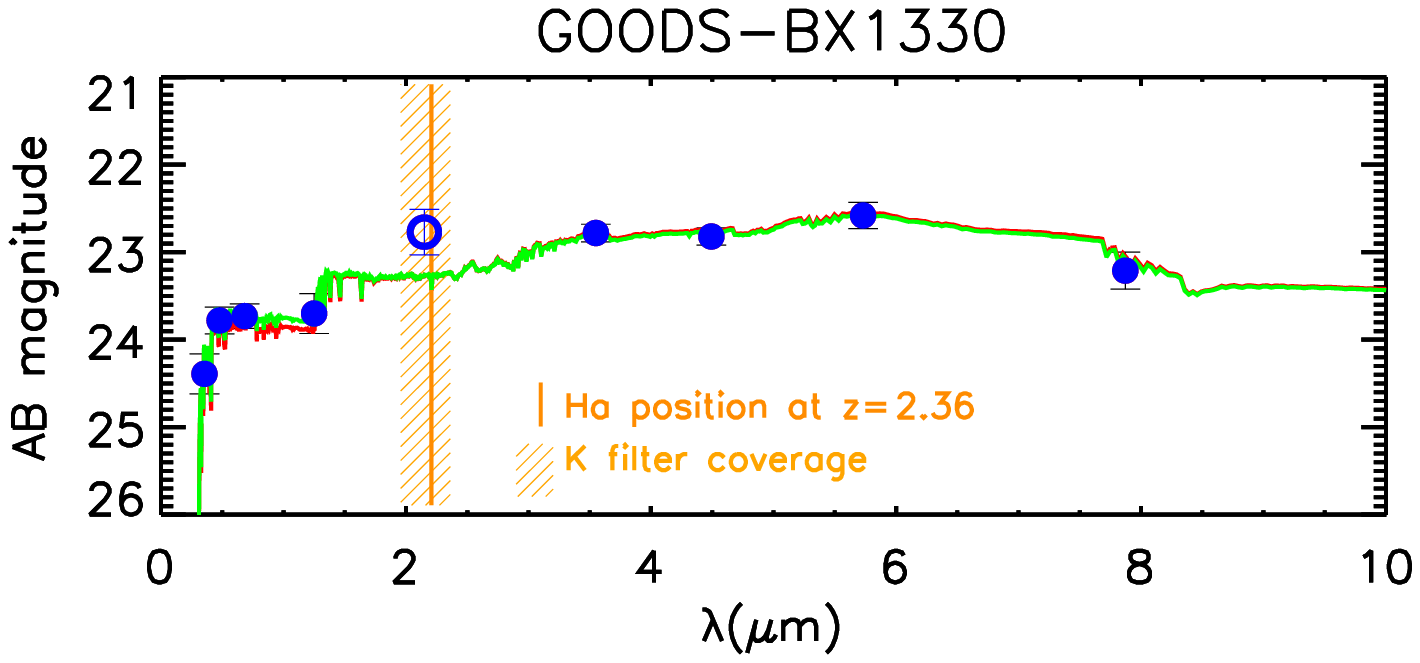}}\\
\subfigure{\includegraphics[trim=0cm 2.8cm 2.4cm 3.3cm,clip=true,width=.45\textwidth]{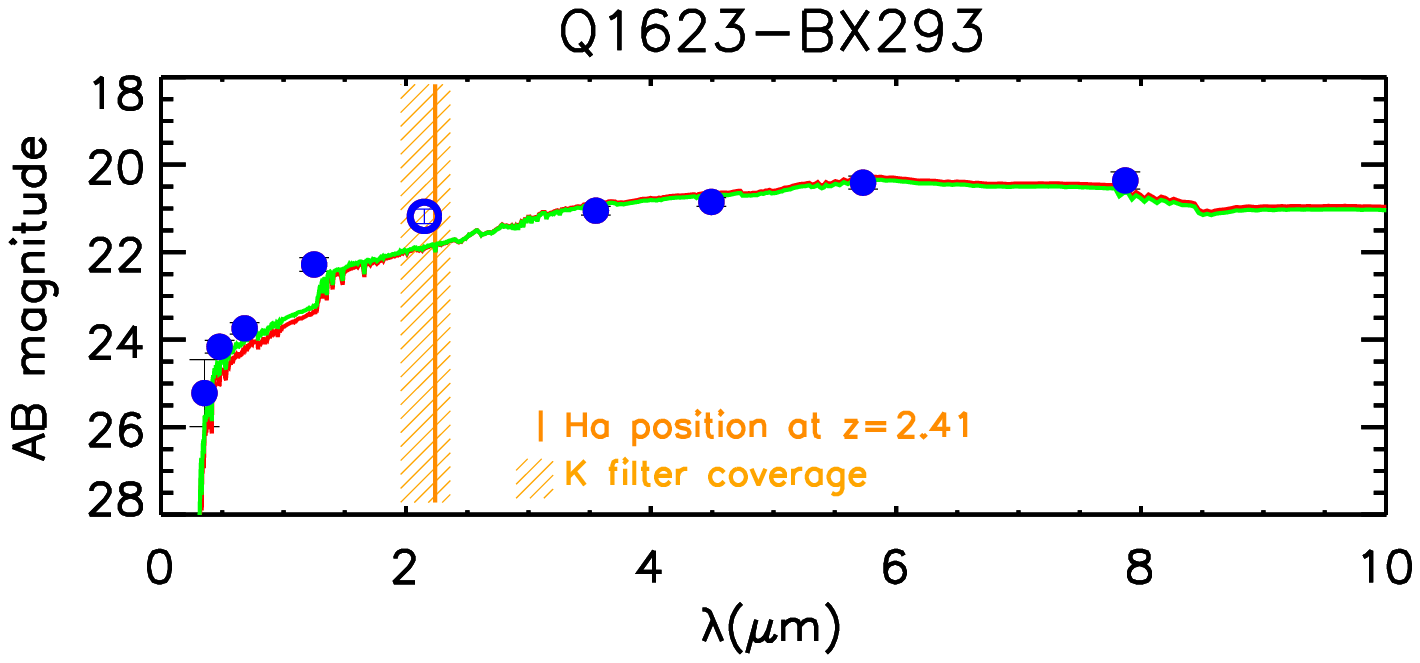}}\\
\subfigure{\includegraphics[trim=0cm 2.8cm 2.4cm 3.3cm,clip=true,width=.45\textwidth]{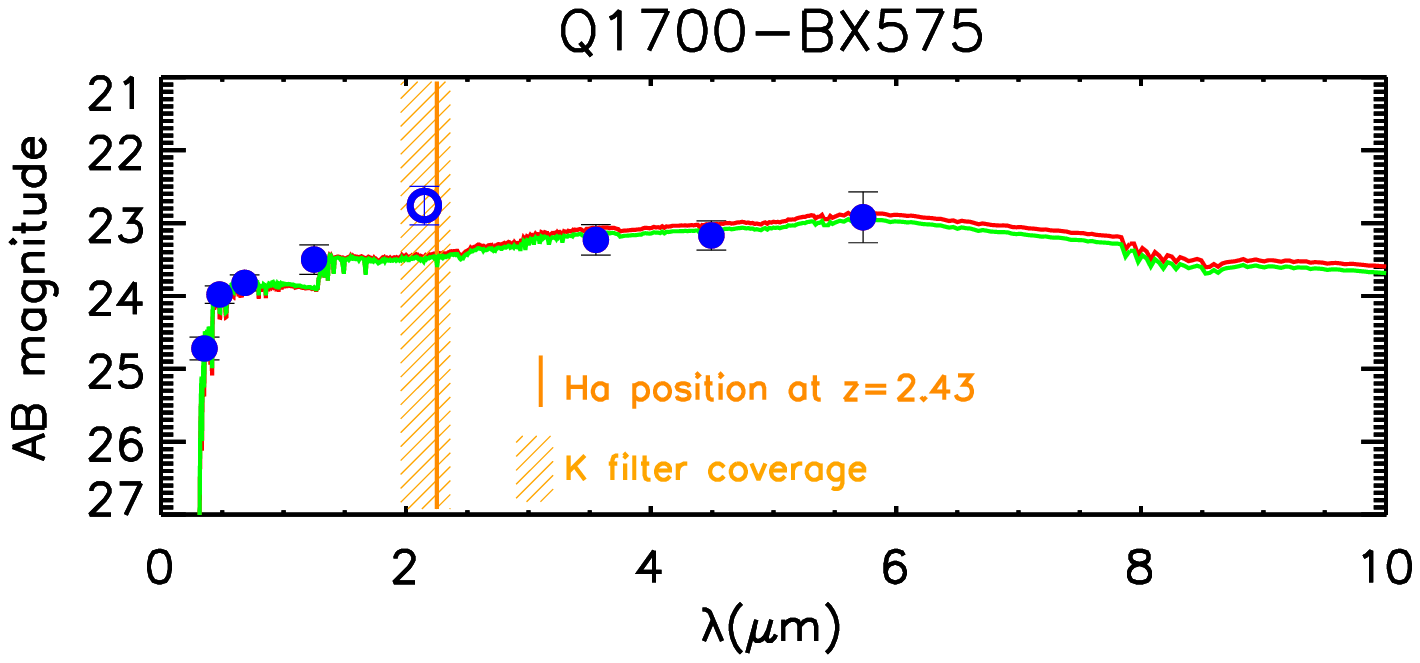}}
\caption{ Best-fit SED models for three objects in our sample: BX1330 in GOODS-North (top), BX293 in Q1623 (middle), and BX575 in Q1700 (bottom). All the models assume an exponentially rising star-formation history, with a lower limit in age of 50\,Myr. The {\em$K_{\rm s}$} band data are shown with an open circle in each plot. The green and red lines show the best-fit models when we exclude and include, respectively, the {\em K}-band photometry in the fitting.
The two fits are similar because the {\em$K_{\rm s}$} photometry is down-weighted in the fitting due to larger measurement errors.
The vertical orange line shows the position of the redshifted {\halpha} line. The dashed area represents the {\em$K_{\rm s}$} filter coverage.
At redshifts of $z\sim 2.08-2.50$, the {\em$K_{\rm s}$} filter is affected by the {\halpha} emission line, and the difference between the $K_{\rm s}$-band flux and the underlying
stellar population model can be used to estimate the line flux. }
\label{fig:sed}
\end{figure}

\section{{\halpha} and UV Luminosity}
\label{sec:sec3}

\subsection{{\halpha} Fluxes and Equivalent Widths}
\label{sec:haflux}

Using the procedure discussed in \S\ref{sec:sedfit}, we fit all the photometry, excluding
{\em$K_{\rm s}$}-band, in order to determine the continuum level at 6564\,\AA.
The continuum magnitude is calculated by multiplying the best-fit SED
model by the {\em$K_{\rm s}$} filter transmission curve.  The
difference between the observed {\em$K_{\rm s}$}-band magnitude and the
continuum magnitude is used to extract the {\halpha} line flux as
follows.  The observed {\em$K_{\rm s}$} magnitude is considered as the sum of
the continuum and the {\halpha} fluxes, while the SED-inferred {\em$K_{\rm s}$} magnitude represents only the continuum flux. Assuming a Gaussian form for the redshifted
{\halpha} line, the change between these fluxes will yield the
{\halpha} flux (Figure~\ref{fig:sed}):

\begin{align}
\frac{\int \frac{d\nu}{\nu}~f_{\halpha}(\nu)~T_K(\nu)}{\int \frac{d\nu}{\nu}~T_K(\nu)} = 10^{\frac{(m_{obs}+48.6)}{-2.5}} - 10^{\frac{(m_{SED}+48.6)}{-2.5}}.
\label{equ:ha}
\end{align}

Here, $f_{\halpha}(\nu)$ is the flux density of the {\halpha} line in
units of 
erg\,s$^{-1}$\,cm$^{-2}$\,Hz$^{-1}$,
$T_\text{K}(\nu)$ is the {\em$K_{\rm s}$} filter transmission curve,
and $m_{obs}$ and $m_{SED}$ are the observed and the continuum
magnitudes, respectively. The {\halpha} flux is corrected for
contamination from the [N{\scriptsize II}] line based on the mass-dependent
[N{\scriptsize II}]-to-{\halpha} flux ratios of \citet{erb06a}. The stellar masses are determined from the SED models and the corresponding [N{\scriptsize II}] line contamination, as listed in Table~\ref{tab:nii}, is used to correct the {\halpha} flux. The SFRs are then calculated using the
\citet{kennicutt98} relation to convert the {\halpha} 
line luminosity to an SFR. 

There are three main sources of uncertainty in the derived
  {\halpha} fluxes. The largest uncertainty is the photometric
  error. 
The typical error in the observed {\em$K_{\rm s}$}-band
  magnitude is $\simeq 0.27$. The second source of uncertainty
  is the [N{\scriptsize II}] correction. 
  The uncertainty on the \citet{erb06a} [N{\scriptsize
      II}]-to-{\halpha} line ratios (see Table~\ref{tab:nii}) is added
  in quadrature to the photometric error. The third source of
  uncertainty is the error associated with the continuum flux. In order to account for this uncertainty, we estimated
  the continuum flux at 6564\AA~ from the
  best-fit model assuming six different star-formation histories:
  exponentially rising, exponentially declining, and constant, for
  each considering all ages and ages greater than 50\,Myr. For 94\% of
  the galaxies the difference in the mean of the continuum fluxes assuming different
star-formation histories to those assuming a 
  rising star-formation history with ages $\ge$ 50\,Myr is less than 0.1 magnitude. 
The average error in the
  estimated continuum magnitude is $\sim 0.05$\,mag, which is
  negligible compared to the observed {\em$K_{\rm s}$}-band magnitude
errors. Combining the three sources of uncertainty discussed above yields
a typical relative error in {\halpha} flux of $\sim 0.29$.

\begin{deluxetable}{cc}
\setlength{\tabcolsep}{0.05in} 
\tabletypesize{\footnotesize}
\tablewidth{0pc}
\tablecaption{{\halpha} Flux Corrections for [N{\scriptsize II}] Contamination\tablenotemark{a}}
\tablehead{
\colhead{ Mass Range ($10^{10} \msun$)\tablenotemark{b}} &
\colhead{ N2\tablenotemark{c}   }
}
\startdata
$< 0.88$ &  $< -1.22$\\
0.88 -- 2.00 & $-1.00^{+0.07}_{-0.09}$\\
2.00 -- 3.69 &  $-0.85^{+0.05}_{-0.06}$\\
3.69 -- 6.03  &  $-0.78^{+0.05}_{-0.05}$\\
6.03 -- 8.82 & $-0.66^{+0.03}_{-0.04}$\\
$> 8.82$ & $-0.56^{+0.02}_{-0.02}$
\enddata
\tablenotetext{a}{Based on \citet{erb06a}}
\tablenotetext{b}{Assuming a \citet{salpeter55} IMF}
\tablenotetext{c}{$N2\equiv\log(\frac{F(\text{[N{\scriptsize II}]})}{F({\text{\halpha})}})$}
\label{tab:nii}
\end{deluxetable}

The equivalent width of the {\halpha} line is estimated by dividing the {\halpha} flux derived from Equation~\ref{equ:ha} by the continuum flux density at the wavelength of the {\halpha} line:

\begin{eqnarray}
(1+z) EW_0 = \frac{F_{\halpha}}{f_{\lambda}^{cont}}.
\end{eqnarray}

The $f_{\lambda}^{cont}$ is the continuum flux density in units of erg~s$^{-1}$~cm$^{-2}$~\AA$^{-1}$ that is estimated through the best-fit SED model, $F_{\halpha}$ is the {\halpha} line flux in erg~s$^{-1}$~cm$^{-2}$, $z$ is the redshift of the galaxy, and $EW_0$ is the rest-frame equivalent width in \AA.

We define whether a galaxy has a ``detected'' {\halpha} line according to the
following.
Galaxies whose {\em$K_{\rm s}$}-band photometry exceeds the continuum
level by more than the {\em$K_{\rm s}$}-band magnitude error are
referred to as ``{\halpha} detections''. Galaxies where the
  {\em$K_{\rm s}$}-band photometry is consistent with the continuum
  level to within $1$\,$\sigma$ are referred to as ``{\halpha}
  non-detections'', and an upper limit of $1\sigma$ above the measured K photometry 
  is used for these objects. Out of 262 galaxies in
  our sample, 149 are detections and 94 are non-detections. There are
  19 objects with {\em$K_{\rm s}$} magnitudes fainter than the
  continuum by more than $1\sigma$. We removed these galaxies from our
  discussion due to their {\em$K_{\rm s}$} photometry being
  inconsistent with the photometry from adjacent bands.

\begin{figure}[tbp]
\includegraphics[width=0.5\textwidth] {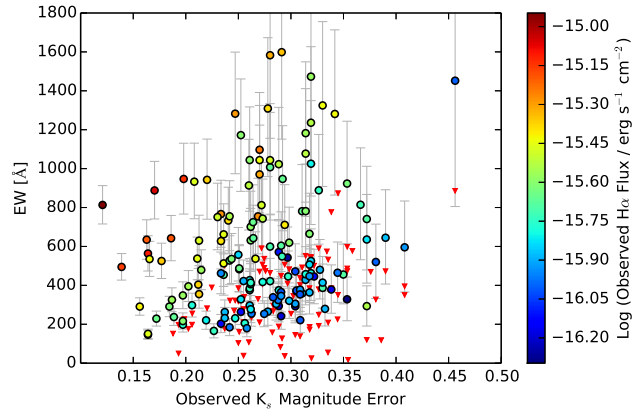}
\caption{
  Estimated {\halpha} rest-frame equivalent widths as a
  function of the error of the observed {\em$K_{\rm s}$}-band
  magnitude. The circles are detected galaxies and are color coded based on their observed {\halpha} fluxes (uncorrected for extinction). The red triangles show $1\sigma$ upper limits.  Detecting the {\halpha} emission requires that the observed {\em$K_{\rm s}$} magnitude lies more that $1\sigma$ above the underlying continuum, where $1\sigma$ is defined by the {\em$K_{\rm s}$} magnitude error. 
  The {\em$K_{\rm s}$} magnitude includes the light both from the {\halpha} emission line and the brightness of the continuum at 6564 \AA. There is a tight correlation between the continuum brightness and the {\em$K_{\rm s}$} magnitude error; the fainter continuum results in larger uncertainty in the observed magnitude.
As a result, there are more undetected objects with large {\em$K_{\rm s}$}-band errors. On the other hand, the brightest {\halpha} objects have bright continuum and small {\em$K_{\rm s}$}-band uncertainties. The deficiency of objects in the upper-left corner is mainly because very bright {\halpha} objects that also have bright continuum are rare.}
\label{fig:ew-omagerr}
\end{figure}

As discussed in \S\ref{sec:intro}, our method of computing {\halpha}
fluxes and EWs has the advantage of being immune to slit-loss
corrections. Our sample covers a wide range of {\halpha} fluxes and
EWs. Figure~\ref{fig:ew-omagerr} shows the distribution of
  {\halpha} EWs and fluxes as a function of the {\em$K_{\rm s}$}
  magnitude $1\sigma$ uncertainty.  The detection of {\halpha} with
  the method adopted here depends on both the {\halpha} line flux and
  the brightness of the continuum at 6564 \AA. As we have defined detections
to be those objects where the {\em$K_{\rm s}$} magnitude exceeds the
continuum level by more than $1\sigma$ , the number of
  undetected objects increases for objects that are fainter in the continuum (Figure~\ref{fig:ew-omagerr}).

\begin{figure}[tbp]
\includegraphics[width=0.5\textwidth] {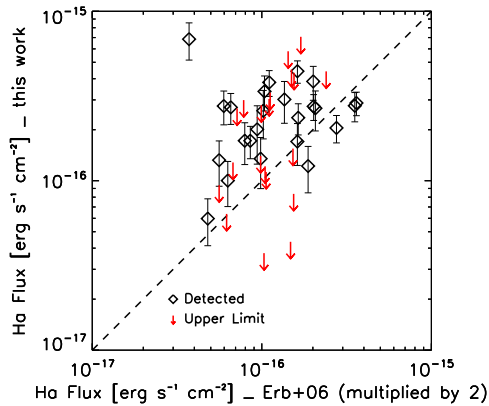}
\caption{A comparison of observed {\halpha} fluxes of the objects in
  common between our sample and that of \citet{erb06c}. The diamonds
  are the detected objects and downward arrows show $1\,\sigma$ upper
  limits, as discussed in the text. The fluxes are measured based on
  the difference between the observed and the continuum {\em$K_{\rm
      s}$} magnitudes predicted from the best-fit model. The error
  bars are dominated by the uncertainty in the observed {\em$K_{\rm
      s}$} photometry. The \citet{erb06c} spectroscopic fluxes are
  multiplied by a factor of 2 to account for slit losses.  The
  dashed line denotes the one-to-one relationship.}
\label{fig:fluxcompare}
\end{figure}

The {\halpha} observed fluxes of 44 objects in common with the
\citet{erb06c} spectroscopic sample are compared in
Figure~\ref{fig:fluxcompare}. The \citet{erb06c} spectroscopic fluxes
are multiplied by a factor of two to account for slit losses and the
aperture used to extract the spectra \citep{erb06c,reddy10}.  The slit loss
correction factor applied to the spectroscopic fluxes depends on
various factors such as the accuracy of the astrometry, the size of
the object convolved with the seeing at the time of observation in
comparison with the size of the slit, and the accuracy of the mask
alignment during the observation.  As noted in \citet{erb06c}, the
factor of two slit-loss correction is an average estimate and
the slit losses will of course vary from object to object, likely
accounting for some of the scatter in Figure~\ref{fig:fluxcompare}.
On the other hand, while broadband photometry has its advantages, our
{\em$K_{\rm s}$}-band measurements suffer from a larger measurement
(random) uncertainty relative to the other methods given the depth of
our ground-based images. Although most of the detected {\halpha}
  fluxes in this study (the black symbols in
  Figure~\ref{fig:fluxcompare}) are higher that the spectroscopic
  fluxes, about half of the sample is estimated as upper limits (the
  red arrows). Considering both the detections and non-detections, the
  broad-band estimated fluxes are generally consistent with the
  spectroscopic fluxes. To quantify the
  degree of the correlation we conduct a generalized Kendall's tau
  ($\tau$) statistic.  The generalized Kendall's $\tau$ rank
  correlation coefficient is a non-parametric test that permits
  non-detections in both variables (here, our estimated {\halpha}
  fluxes).  Kendall's $\tau$ correlation coefficient for our flux
estimation with the \citet{erb06c} measurements, excluding the one
outlier object, is $\tau$ = 0.17, with significance of p-value = 0.11.

\subsection{UV Luminosities and SFRs}
\label{sec:uvlum}
 
UV luminosities are calculated using the fluxes of the best-fit SED
models at 1700\,\AA. The \citet{kennicutt98} relation to convert the
UV luminosity (over the wavelength range 1500-2800\,\AA\,) to SFR
applies only to galaxies where star formation proceeds for $\ga 100$\,Myr.
However, there are galaxies in our sample with inferred ages between 50
and 100\,Myr.
For these younger galaxies, we used an age- and
star-formation-history-dependent relation in the conversion of the UV
luminosity to SFR, as discussed in \citet{reddy12b}. The age-dependent
conversion resulted in UV SFRs that are at most $14\%$ larger than the
\citet{kennicutt98} SFRs for young galaxies. 
Throughout this analysis, we also use SFRs inferred from the best-fit
SED models. As expected, within the uncertainties, the SED SFRs are
highly correlated with dust corrected UV SFRs. 
  
\section{The IR Luminosity and Bolometric Properties }
\label{sec:irlum}

To estimate the bolometric luminosities of galaxies in our sample,
we infer the infrared luminosity (integrated between 8-1000~$\mu$m) as
discussed below, and add this to the unobscured UV luminosity (i.e.,
observed UV luminosity; \S\ref{sec:uvlum}), as in \citet{reddy10}.

Infrared luminosities, $L_{\rm IR}$, are estimated by using the {\it
  Spitzer}/MIPS 24\,$\mu$m observations.  The $24$\,$\mu$m band is
sensitive to the rest-frame 7.7~$\mu$m PAH emission, which correlates
with $L_{\rm IR}$ \citep[e.g.,][]{chary01,dale02,elbaz11}. To convert observed 24\,$\mu$m
magnitudes to total IR luminosity (L$_{IR}$), we used several dust SED
templates, including those of \citet{chary01}, \citet{dale02}, and
\citet{rieke09}. Luminosities determined from the rest-frame 8~$\mu$m
flux density alone, regardless of the dust template used, tend to over
predict L(IR), particularly for LIRGs \citep[e.g.,][]{reddy12a} and
ULIRGs \citep[e.g.,][]{nordon10,magnelli11}.  In order to account for
  the luminosity-dependent overestimation of the derived L(IR), we use
  a correction equation described in \citet{reddy12a}.  The correction
  is calculated by comparing the L(IR) computed from 24, 100,
  160\,$\mu$m, and 1.4\,GHz fluxes, and the L(IR) determined solely
  from 24\,$\mu$m data for a similarly-selected sample of galaxies:
\begin{align}
\label{equ:it-tot}
 \log L_{\text{IR}}(\text{all data}) =&\, 0.537 \times \log L_{\text{IR}}(\text{24}\mu \text{m}) + 5.136.
\end{align} 
All luminosities are in L$_{\odot}$.

Equation\,\ref{equ:it-tot} is based on the \citet{chary01}
models. There is a factor of $\sim 2$ variation in L(IR) derived from
24$\mu$m data using \citet{dale02} and \citet{rieke09} dust templates
\citep[e.g., see][]{reddy12a}. However, once corrected using the
appropriate equations similar to Equation\,\ref{equ:it-tot}, all the
templates result in L(IR) estimates that are consistent with each
other and with the L(IR) computed from 24, 100, 160\,$\mu$m, and
1.4\,GHz fluxes, within the uncertainties of the measured IR
  luminosity as explained in \S\ref{sec:sample-b} \citep[see
    also][]{reddy12a}.  In the subsequent analysis, we use infrared
luminosities inferred from the \citet{chary01} templates.

The IR luminosity is converted to IR (dust-obscured) SFRs using the
\citet{kennicutt98} relation. The sum of IR- and UV-inferred SFRs are
then used to estimate the bolometric SFRs.

\section{Dust Attenuation of the Nebular Regions and Stellar Continuum}
\label{sec:sec5}
\begin{figure*}[!htp]
\centering
\subfigure{\includegraphics[trim=1.1cm 0cm 3.cm 0cm,clip=true,width=0.48\textwidth]{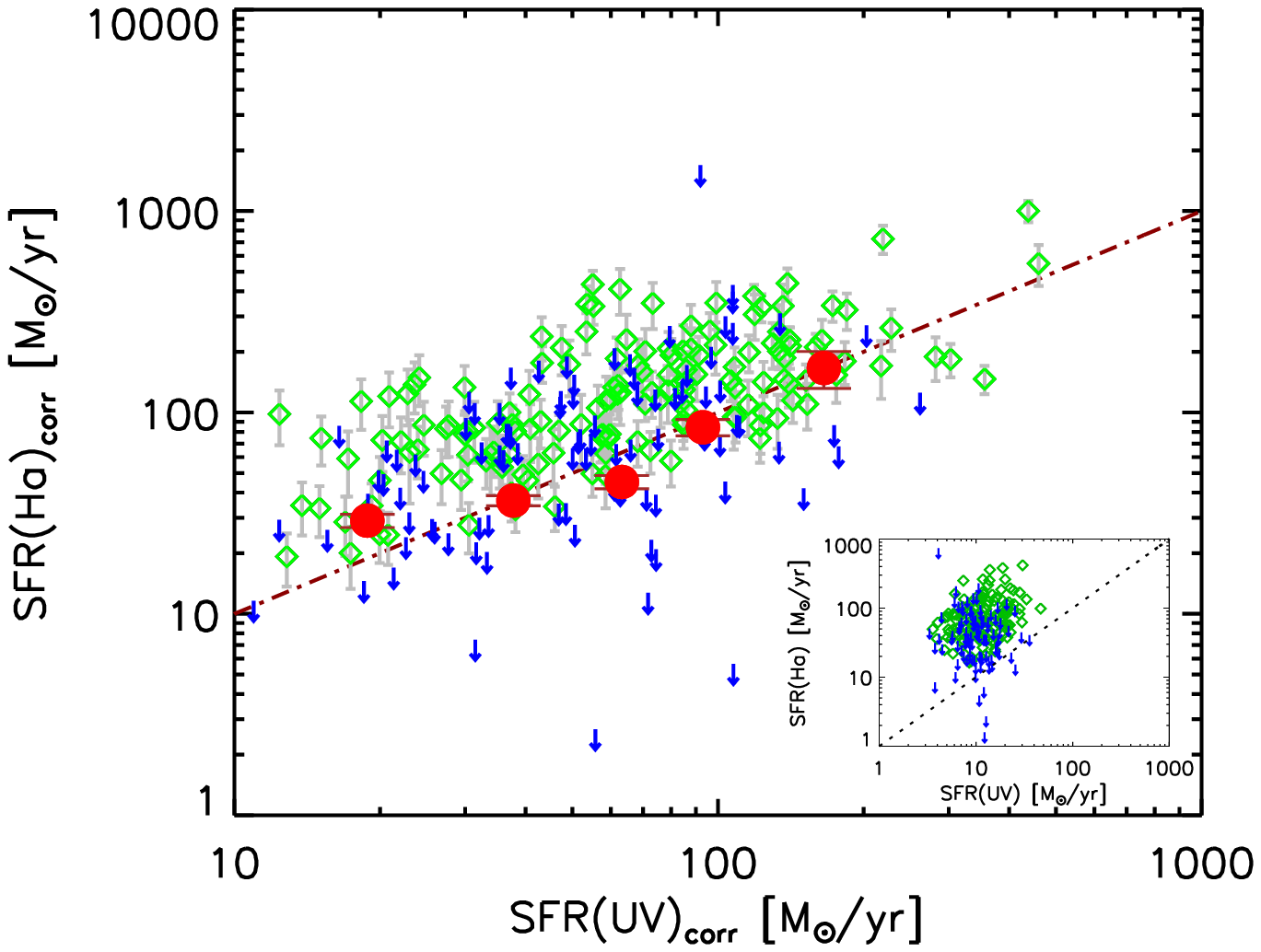}}\quad
\subfigure{\includegraphics[trim=1.1cm 0cm 3.cm 0cm,clip=true,width=0.48\textwidth]{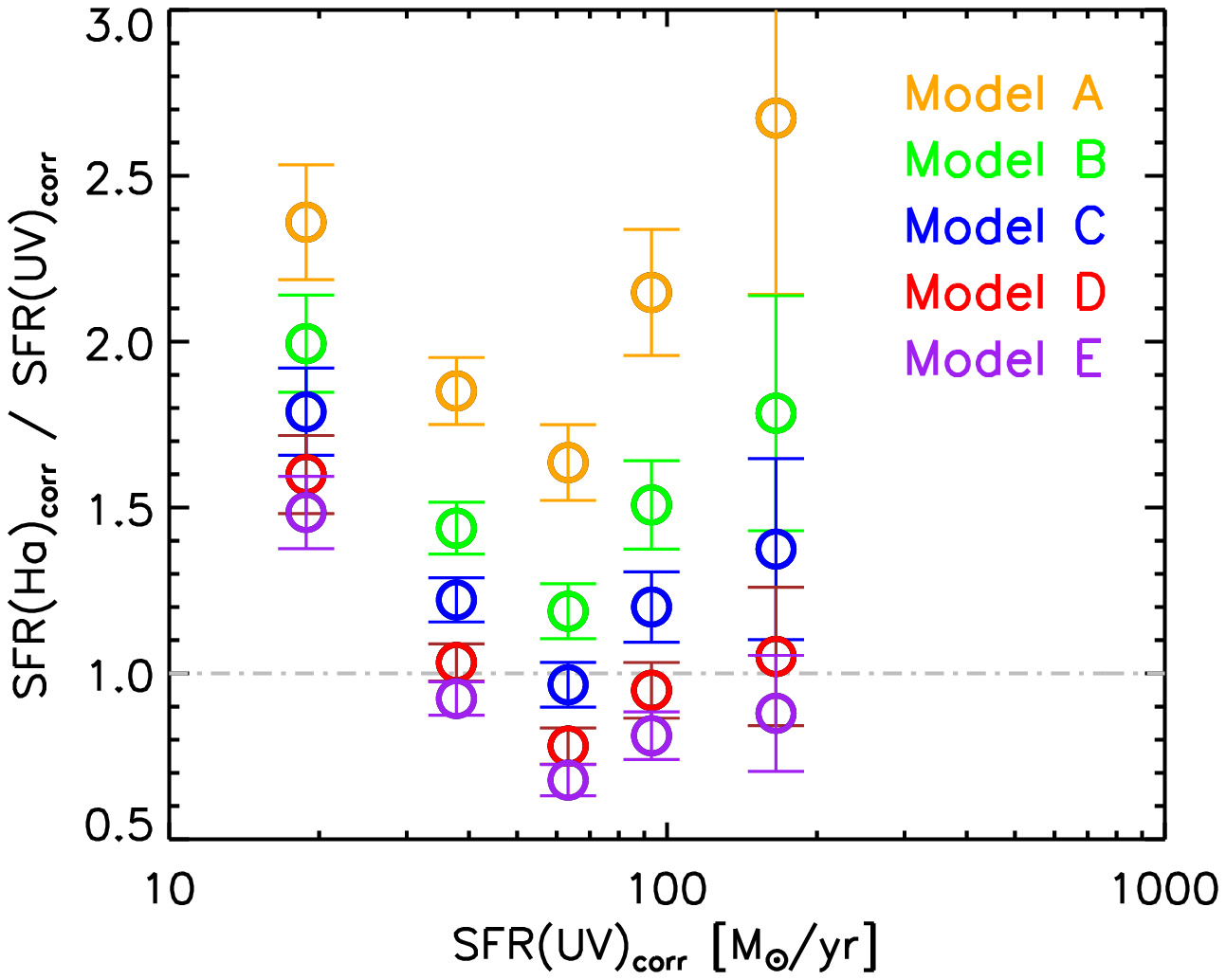}}
\caption{({\em Left:}) Comparison of SFR({\halpha}) and SFR(UV), both
  corrected for dust assuming the Calzetti attenuation curve and
assuming that the same color excess applies for the stellar and nebular
regions. The uncorrected values are shown in the inset panel.
  Green diamonds denote ``detected'' objects, those whose {\em$K_{\rm
      s}$}-band magnitudes are more than 1\,$\sigma$ brighter than the 
  continuum. Blue arrows show the remaining ``undetected'' objects. Red filled
  circles are the results of stacked SEDs in bins of dust-corrected
  SFR(UV), including both the detected and undetected objects.
  Undetected objects bring the stacks lower than 
 the average of the individually detected green diamonds.
The dashed lines indicate the one-to-one relationships. ({\em Right:})
Ratio of SFR({\halpha})$_{corrected}$ to SFR(UV)$_{corrected}$ in bins
of SFR(UV)$_{corrected}$. The colors denote different recipes for the
nebular dust correction. Using the same color excess for nebular
extinction together with the Calzetti curve (model D; see text) gives
the best agreement between {\halpha} and UV SFRs, while assuming
higher nebular attenuation (i.e., $E(B-V)_{nebular} =
2.27\,E(B-V)_{stellar}$) with the Calzetti curve results in
SFR({\halpha}) that is significantly larger than SFR(UV) (model A).
The details of each model are summarized in Table~\ref{tab:sfr}. }
\label{fig:sfrhauv}
\end{figure*}

SFRs inferred from the UV luminosity and the {\halpha} fluxes are
shown in the left panel of Figure~\ref{fig:sfrhauv}. Without dust
corrections, the UV and {\halpha} SFRs have a large discrepancy due to
the smaller dust absorption cross-section at 6564\,\AA\, relative to
that at 1700\,\AA\, (see the inset panel in
Figure~\ref{fig:sfrhauv}). The main left panel of
Figure~\ref{fig:sfrhauv} shows {\halpha} and UV SFRs, both corrected
for dust attenuation based on the Calzetti attenuation curve and
assuming that the same color-excess, $E(B-V)$, as derived from the
best-fit SED model, applies to the stellar continuum and the nebular
regions. To better quantify the mean trend, measurements have also
been performed in bins of UV dust-corrected SFR (filled circles in
Figure~\ref{fig:sfrhauv}). The bins are $\Delta$log(SFR$_{UV,
  corrected}$) = 0.3\,dex wide. For all the galaxies in each bin, 
  regardless of their redshift, we calculate the median flux densities in the observed filters.
  With the new set of median fluxes, the best-fit stellar population model is determined
through $\chi^2$ minimization for the mean redshift of the galaxies in each bin.
  The {\halpha} and UV luminosities are then derived from the
stacked SED in the same way as for individual galaxies (see
\S~\ref{sec:sec3}). Reported errors are calculated based on the
  $1\,\sigma$ standard error of the mean of the {\em$K_{\rm s}$}
magnitudes contributing to each bin.

\begin{deluxetable*}{ccccccc} 
\setlength{\tabcolsep}{0.02in} 
\tabletypesize{\footnotesize}
\tablewidth{0pc}
\tablecaption{Dust Corrected {\halpha} SFRs Using Different Attenuation Recipes, in Bins of SFR(UV)$_{corr}$}
\tablehead{
\colhead{ SFR(UV)$_{corr}$\tablenotemark{a}} &
\colhead{ N\tablenotemark{b}} &
\multicolumn{5}{c}{ SFR({\halpha})$_{corr}$\tablenotemark{c}} \\
\colhead{} &
\colhead{} &
\colhead{Calzetti} &
\colhead{Galactic} &
\colhead{SMC} &
\colhead{Calzetti} &
\colhead{Galactic}\\
\colhead{} &
\colhead{} &
\colhead{Different $E(B-V$)\tablenotemark{d}} &
\colhead{Different $E(B-V)$\tablenotemark{e}} &
\colhead{Different $E(B-V)$\tablenotemark{f}} &
\colhead{Equal $E(B-V)$\tablenotemark{g}} &
\colhead{Equal $E(B-V)$\tablenotemark{h}} \\
\colhead{} &
\colhead{} &
\colhead{(Model A)} &
\colhead{(Model B)} &
\colhead{(Model C)} &
\colhead{(Model D)} &
\colhead{(Model E)}
 }
\startdata
$18\pm 3$ &26& $44\pm 3$ & $38\pm 3$ & $34\pm 2$ & $30\pm 2$ &  $28\pm 2$ \\
$32\pm 6$ &63& $70\pm 4$ & $54\pm3$ & $46\pm 3$ & $38\pm 2$ & $35\pm 2$  \\
$65\pm 12$ &83& $103\pm 7$ & $75\pm 5$ & $61\pm 4$ & $49\pm 3$ & $43\pm 3$  \\
$120\pm 22$  &56& $200\pm 18$ & $140\pm 12$ & $112\pm 10$ & $88\pm 8$ & $76\pm 7$ \\
$258\pm 94$ &15& $443\pm 88$ & $295\pm 59$ & $228\pm 45$ & $174\pm 35$ & $146\pm 29$  
\enddata

\tablenotetext{a}{ The mean and standard deviation of the corrected SFR(UV) in each bin, in units of $\msun\,\text{yr}^{-1}$.}
\tablenotetext{b}{Number of objects in each bin}
\tablenotetext{c}{ The SFR({\halpha})$_{corr}$ is derived based on the stacked SEDs in bins of dust-corrected SFR(UV), in units of $\msun\,\text{yr}^{-1}$. The errors are $1\,\sigma$ standard error of the mean of the {\em$K_{\rm s}$} magnitudes contributing to each bin, which is converted to the error in SFR.}
\tablenotetext{d}{ $\text{A}_{\halpha} = 2.27~\kappa_{\raisebox{-1pt}{\tiny{\text{Calz}}}}({\halpha})~E(B-V)_{\text{stellar}}$~; $\text{A}_{\halpha}$ is the total nebular extinction at 6564 \AA ~and $\kappa_{\raisebox{-1pt}{\tiny{\text{Calz}}}}({\halpha})$ is the Calzetti reddening at 6564 \AA.}
\tablenotetext{e}{$\text{A}_{\halpha} = 2.27~\kappa_{\raisebox{-1pt}{\tiny{\text{Gal}}}}({\halpha})~E(B-V)_{\text{stellar}}$~; $\kappa_{\raisebox{-1pt}{\tiny{\text{Gal}}}}({\halpha})$ is the Cardelli Galactic reddening at 6564 \AA.}
\tablenotetext{f}{$\text{A}_{\halpha} = 2.27~\kappa_{\raisebox{-1pt}{\tiny{\text{SMC}}}}({\halpha})~E(B-V)_{\text{stellar}}$~; $\kappa_{\raisebox{-1pt}{\tiny{\text{SMC}}}}({\halpha})$ is an SMC reddening at 6564 \AA.}
\tablenotetext{g}{$\text{A}_{\halpha} = ~\kappa_{\raisebox{-1pt}{\tiny{\text{Calz}}}}({\halpha})~E(B-V)_{\text{stellar}}$~; $\kappa_{\raisebox{-1pt}{\tiny{\text{Calz}}}}({\halpha})$ is the Calzetti reddening at 6564 \AA.}
\tablenotetext{h}{$\text{A}_{\halpha} = ~\kappa_{\raisebox{-1pt}{\tiny{\text{Gal}}}}({\halpha})~E(B-V)_{\text{stellar}}$~; $\kappa_{\raisebox{-1pt}{\tiny{\text{Gal}}}}({\halpha})$ is the Cardelli Galactic reddening at 6564 \AA.}
\label{tab:sfr}
\end{deluxetable*}

The appropriate dust corrections to apply to the nebular emission lines is still
a subject of debate, as discussed in Section~\ref{sec:intro}.
The attenuation of the nebular lines and the UV stellar continuum are
not completely decoupled as both arise from dust around massive
stars. On the other hand, whether the color-excess is the same for the nebular and stellar regions
\citep[e.g.,][]{erb06c,reddy10} or not
\citep[e.g.,][]{calzetti00,forster09,garn10} in high-redshift galaxies
has yet to be fully investigated.

\citet{calzetti00} found that $E(B-V)_{\rm stel} = 0.44\times E(B-V)_{\rm neb}$ for a sample
of local star-forming galaxies.  In the absence of direct measurements of the nebular
color-excess (e.g., via the Balmer decrement), this relationship can in principle be used
to estimate the nebular color excess and apply a dust correction to the {\halpha} line; 
e.g., 
\begin{align}
\label{diffatt}
\text{A}_{\halpha} &= \kappa_{\raisebox{-1pt}{\tiny{\text{Gal}}}}({\halpha})~E(B-V)_{\text{nebular}}\nonumber\\
&= 2.27~\kappa_{\raisebox{-1pt}{\tiny{\text{Gal}}}}({\halpha})~E(B-V)_{\text{stellar}}\nonumber\\
&= 2.27\times2.52~E(B-V)_{\text{stellar}}\nonumber\\
&= 5.72~E(B-V)_{\text{stellar}},
\end{align}
where $\kappa_{\raisebox{-1pt}{\tiny{\text{Gal}}}}$ is the
\citet{cardelli89} Galactic extinction curve, assuming that the
  nebular regions in the high-redshift galaxies abide by such a dust
  curve. $A_{\halpha}$ is the absolute extinction of the {\halpha} emission line, and $E(B-V)_{\rm stellar}$ is the SED-inferred stellar color excess.
 
We investigate several of the more commonly used dust-correction recipes,
as described below.  The {\halpha} SFRs are
corrected using five different methods:
\renewcommand{\theenumi}{\Alph{enumi}}
\begin{enumerate}
\item the Calzetti attenuation curve for both gas and stars, but 2.27$\times$ larger color-excess for the nebular lines than the stellar continuum (as we call it ``different $E(B-V)$''), 
\item the Calzetti attenuation curve for the stellar continuum, and the \citet{cardelli89} Galactic curve for the nebular lines assuming 2.27$\times$ larger color-excess, 
\item the Calzetti attenuation curve for the stellar continuum, and an SMC extinction curve \citep{gordon03} for the nebular lines assuming 2.27$\times$ larger color-excess, 
\item the Calzetti attenuation curve for both gas and stars, and the same color-excess (as we call it ``equal $E(B-V)$''), and 
\item the Calzetti attenuation curve for the stellar continuum, an SMC extinction curve \citep{gordon03} for the nebular lines assuming the same color-excess.
\end{enumerate}
We choose to compare these different dust-correction scenarios for the
stacked values instead of the individual galaxies as the former
represent an average over many points including both detected and
undetected quantities, each of which may be relatively uncertain.  The
results are plotted in the right panel of Figure~\ref{fig:sfrhauv} and
are reported in Table~\ref{tab:sfr}.

To summarize, these are the values that are used to correct the observed {\halpha} SFRs in each model according to Equation~\ref{diffatt}:
\begin{align}
\text{A}_{\halpha}^{(A)} &= 2.27~\kappa_{\raisebox{-1pt}{\tiny{\text{Calz}}}}({\halpha})~E(B-V) = 7.56~E(B-V), \nonumber \\
\text{A}_{\halpha}^{(B)} &= 2.27~\kappa_{\raisebox{-1pt}{\tiny{\text{Gal}}}}({\halpha})~E(B-V) = 5.72~E(B-V), \nonumber\\
\text{A}_{\halpha}^{(C)} &= 2.27~\kappa_{\raisebox{-1pt}{\tiny{\text{SMC}}}}({\halpha})~E(B-V) = 4.54~E(B-V), \nonumber\\
\text{A}_{\halpha}^{(D)}&= \kappa_{\raisebox{-1pt}{\tiny{\text{Calz}}}}({\halpha})~E(B-V) = 3.33~E(B-V), \nonumber \\
\text{A}_{\halpha}^{(E)}&= \kappa_{\raisebox{-1pt}{\tiny{\text{Gal}}}}({\halpha})~E(B-V) = 2.52~E(B-V). 
\end{align}
 In these equations, $E(B-V)$ is the SED-inferred color-excess observed for the stellar continuum and A$_{\halpha}$ is the total nebular extinction at 6564 \AA.

In model (A) with a larger color excess for the nebular regions with the Calzetti curve, {\halpha}-inferred SFRs are significantly larger than UV SFRs. Taking the average of the SFR({\halpha}) (model A) to the corrected SFR(UV) in the five bins of SFR(UV) indicated in Tables~\ref{tab:sfr} yields a value of $2.1\pm 0.2$ which shows a $\sim 5\sigma$ discrepancy from unity.

Using either model (C) or (D), results in SFR$_{corrected}({\halpha})$ that are in good agreement with SFR$_{corrected}(UV)$. Model (B) reproduce systematically higher SFRs but is still consistent with the range of SFR$_{corrected}(UV)$ in higher bins, and model (E) estimations are lower than expected.

 Based on the data provided in Table~\ref{tab:sfr} and the right panel
 of Figure~\ref{fig:sfrhauv}, model (D) provides the best agreement
 between SFRs among the five models. The average of the SFR({\halpha})-to-SFR(UV) ratios in the five SFR(UV) bins of this model shows less than $1\sigma$ discrepancy from unity.
 This suggests that {\em on average}
 the best recipe to correct the {\halpha}-inferred quantities of star-forming galaxies at 
 $z\sim 2$ is to use $E(B-V)_{\text{nebular}} =
 E(B-V)_{\text{stellar}}$ with the Calzetti reddening curve.

\begin{figure}[tbp]
\includegraphics[trim=2.5cm 0cm 0cm 0cm,clip=false,width=.65\textwidth]{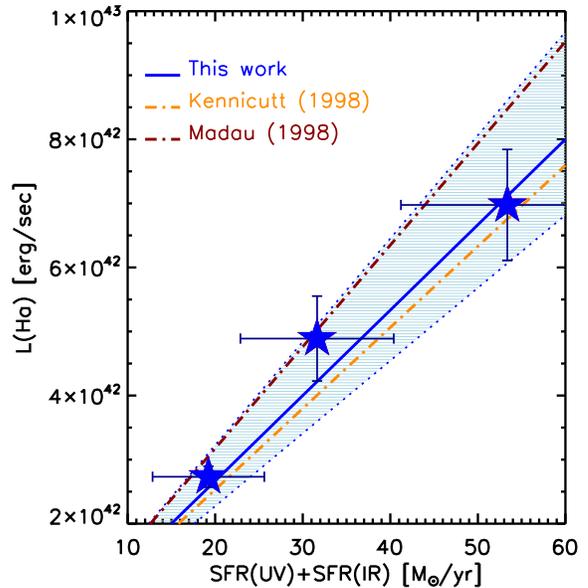}
\caption{ The observed {\halpha} line luminosity as a function of
  bolometric SFR. The stars indicate the average values in bins of
  SFR(SED).  Galaxies with SFR(SED) $<$ 32 $\msun$~yr$^{-1}$ are in
  the first bin; those with SFR(SED) between 32 $\msun$~yr$^{-1}$ and 100
  $\msun$~yr$^{-1}$ are in the middle bin; and last bin consists of
  galaxies with SFR(SED) $>$ 100 $\msun$~yr$^{-1}$. The error in SFR(UV)+SFR(IR)
  is determined by the dispersion error in IR luminosity of the objects contributing 
  to each stack. L(IR) dispersion error is estimated through bootstrap resampling
  simulations. The solid blue line is the best linear fit. The
  slope of this fit gives us the conversion factor for
  L({\halpha})$_{obs}$-to-SFR$_{total}$ relation. The shaded area is
  the 68\% confidence interval of our estimated slope. The orange and
  dark-red dashed lines show the conversion factors of
  \citet{kennicutt98} and \citet{madau98}, respectively.}
\label{fig:lha}
\end{figure}

\section{Bolometric Star-Formation Rates}
\label{sec:sec6}
 
As discussed above, a primary disadvantage of using {\halpha} and UV
SFR diagnostics is that we must account for dust extinction, though
this is less of a problem for {\halpha} as it is for the UV. The
attenuation curves (e.g., the \citealt{calzetti00} curve) that are
used to correct the luminosities encode information regarding the dust
grain size distribution and the geometrical distribution of dust with
respect to stars (e.g., whether the dust is clumpy or uniform, a
homogeneous mixture or a foreground screen, etc.).  On the other hand,
the IR luminosity is a direct tracer of dust and thus provides an
independent and more robust diagnostic of dust attenuation.\footnote{
  The dust heated by the visible light from older stellar populations
  (i.e., the cold component) also contributes to the IR
  luminosity. However, in case of the star-forming galaxies the
  contribution of the UV radiation of massive stars (the warm
  component) is expected to dominate the total IR luminosity
  \citep{kennicutt98}.}  Adding the IR luminosity, which accounts for
the obscured star formation, to the unobscured (UV and {\halpha})
tracers of SFR can give a more reliable estimate of the bolometric
SFR.

\subsection{The Relationship between Observed {\halpha} Luminosity and Total SFR}
\label{sec:sfrbol}

To convert the observed {\halpha} luminosity to total SFR, the most
commonly used calibration is that of \citet{kennicutt98}, which is
computed using an evolutionary synthesis model, assuming Case B
recombination with $T_e$=10,000\,K.  In the absence of Balmer
decrement measurements, the observed H$\alpha$ SFR must be corrected
for dust with some assumption of the attenuation curve and
color-excess of the ionized gas.  In this section, we examine the
L$_{\halpha}$-to-SFR conversion by relying only on the observed data,
without making any assumptions about the dust correction or electron
temperature of the ionized gas. With this conversion, one can use the
{\halpha} observed luminosity as a proxy for total SFR without
assuming a dust attenuation curve and its associated
uncertainties. The conversion applies to stellar populations with
constant star formation at least over 100\,Myr.  For bursty star
formation histories over timescales shorter than 100\,Myr, the
{\halpha} and UV+IR luminosities diverge as the {\halpha} luminosity
traces more instantaneous star formation over timescales of $\sim
10$\,Myr, while IR+UV is not sensitive to the change of star formation
on timescales $\lesssim 100$\,Myr. Here, we are deriving the
conversion factor based on the stacked data that represents the
``average'' quantities and thus should be used with caution for
galaxies that may be undergoing bursty star formation.

Bolometric SFR (SFR(IR)+SFR(UV)) is plotted as a function of the
{\halpha} luminosity in Figure~\ref{fig:lha}.  As both the
{\halpha} and IR measurements have non-detections, we decided to bin
the data with respect to the SED-inferred SFR. We chose the bins so that
in each bin the SFR(SED) is consistent with the stacked
SFR(IR)+SFR(UV) within the uncertainties, because ultimately we are using SFR(IR)+SFR(UV) to calculate the {\halpha} luminosity-to-SFR conversion factor. MIPS data are
stacked in each bin and the total IR luminosity is extracted in the
same way as for individual galaxies. The UV and {\halpha} luminosities
are derived from the stacked SEDs.

A weighted least squares regression method is used to find the
slope of the relation:

\begin{align}
\label{equ:sfr-lum}
\text{SFR}_{bol} \,(\msun \, \text{yr}^{-1}) = (7.5 \pm 1.3) \times10^{-42}\, \text{L}_{obs}(\halpha) \, (\text{erg} \, \text{s}^{-1}).
\end{align}

This relation can be used to convert the observed {\halpha} luminosity
in units of erg\,s$^{-1}$ to the bolometric SFR. 
The factor given in \citet{kennicutt98} and \citet{kennicutt94}, to
convert observed/intrinsic luminosity to observed/intrinsic SFR, is
7.9 $\times10^{-42}$  $\msun$~yr$^{-1}$~erg$^{-1}$~s. \citet{madau98} reported a
conversion factor of 6.3 $\times10^{-42}$ $\msun$~yr$^{-1}$~erg$^{-1}$~s. 
These factors
are computed using evolutionary synthesis models and are subject to
those models' uncertainties as well as the initial assumptions that
went into these models, such as IMF, star-formation history, and the stellar evolution and atmosphere models. For the same galaxy type and assumed IMF, \citet{kennicutt98} reports $\sim 30\%$ variation among different calibrations, which mainly reflects the sensitivity to the SED modeling stellar evolution input data.
In our analysis, converting L(IR) and L(UV) to SFR still rely on the
model assumptions, but previous studies have shown that in the
absence of the Balmer decrement measurements, the sum of SFR(IR) and
SFR(UV) is the most reliable estimate of total SFR \citep{hirashita03,reddy12b}.
Our derived slope for a Salpeter IMF is consistent with \citet{kennicutt98} and \citet{madau98} within the errors.

Excluding galaxies with ages $<$ 100\,Myr (see \S~\ref{sec:sfrsed}) 
results in a best-fit slope of 7.8 $\times10^{-42} \msun$~yr$^{-1}$~erg$^{-1}$~s that is consistent with the slope found above.

\begin{figure}[tbp]
\includegraphics[trim=2.5cm 0cm 0cm 0cm,clip=false,width=.65\textwidth]{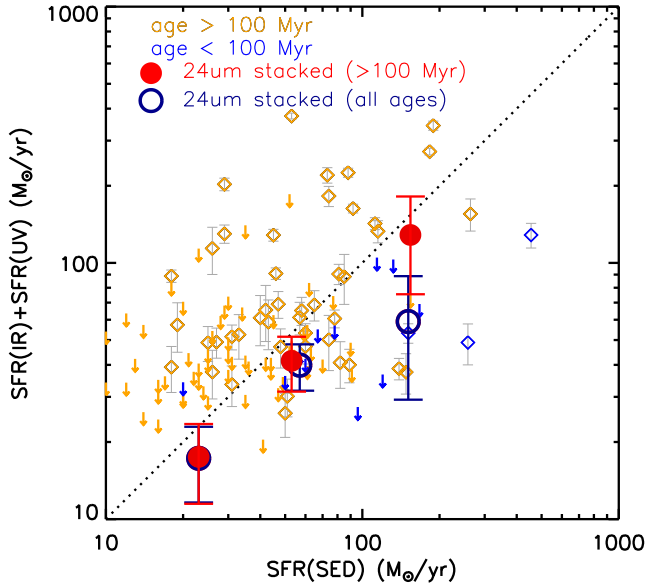}
\caption{Comparison of the bolometric
  SFR and SED-inferred SFR.  The values are divided into three bins of
  SFR(SED) including individual detections and individually undetected galaxies at 24\,$\mu$m; Open blue circles include all objects and filled red circles
  include only those galaxies older than 100\,Myr.
  Diamonds are individual detected objects and
  downward arrows represent $3\sigma$ upper limits in SFR(IR)+SFR(UV) for
the MIPS undetected galaxies. Young galaxies (ages $<$ 100\,Myr) are colored blue
  and galaxies with ages $>$ 100\,Myr are orange. SFR(IR) error bars
  are estimated through bootstrap simulations and show the dispersion in the SFRs
  contributing to each stack.  The last bin
  consists of galaxies with SFR(SED) $>$ 100 $\msun$~yr$^{-1}$; by removing
  young galaxies (ages $<$ 100\,Myr) about half of the sample in this bin is
  dismissed and we see the change between the two measurements (including
  and excluding the young ones).
 The dotted line indicates a one-to-one correspondence.}
\label{fig:sed_bol}
\end{figure} 
 
\subsection{Validity of the Calzetti Dust Attenuation for Young Galaxies}
\label{sec:sfrsed}

Figure~\ref{fig:sed_bol} shows the bolometric SFR vs. median SFR
inferred from the best-fit SED model, in bins of SFR(SED).
Both quantities are representative of
the total SFR of the galaxies, the only difference being the method by
which dust is accounted for. The bolometric SFR,
as mentioned before, is the sum of SFR(IR) and SFR(UV), and is
independent of any dust correction.  On the other hand, for SFR(SED),
the effect of dust is considered by using the Calzetti attenuation
curve in the SED fitting procedure. We would like to determine whether for all types of galaxies at $z\sim 2$ the bolometric SFRs agree with SED SFRs that are corrected by the Calzetti attenuation curve.
We stack the MIPS data of individual detected and undetected galaxies in three bins of SFR(SED)
as described in \S\ref{sec:sample-b}. The three bins in log(SFR(SED)) are $<$ 1.5, 1.5 to 2, and $>$ 2, with respectively 44, 56, and 15 galaxies in each bin. SFR(SED)
and SFR(UV) are the median SFRs of the individual galaxies in each
bin. The results of stacking is shown with open blue circles in Figure~\ref{fig:sed_bol}. In the last two bins, the SED SFR is overestimated compared to the bolometric SFR.
Stacking the MIPS images for only objects that are inferred to be
older than 100\,Myr produces different results. The number of objects in bins with galaxies older than 100\,Myr reduces to 43, 50, and 8, respectively from the lowest SFR(SED) bin to the highest.

Once we remove galaxies with ages $<$ 100\,Myr, SFR(SED) agrees well with
SFR(UV)+SFR(IR). An age-dependent L(UV)-to-SFR calibration (as we use here) does not decrease the difference between SFR(SED) and the bolometric SFR for stacks of all galaxies.
It seems younger galaxies are at the same time less
dusty (smaller bolometric SFR) but redder (larger SFR(SED)) than their
older counterparts.
These results are consistent with those of \citet{reddy06b}, \citet{reddy10}, and
\citet{wuyts12}, who find that galaxies with young stellar population
ages are less dusty for a given UV slope than older galaxies. An
SMC-like curve may be more appropriate for these galaxies. One
possibility is that young galaxies have larger dust covering fractions
than older galaxies \citep{reddy10}. A large dust covering fraction
makes the UV slope redder for a given amount of dust attenuation.
This study confirms that at $z\sim 2$ the Calzetti attenuation curve is applicable to star-forming galaxies older than 100\,Myr, but a steeper attenuation curve may be necessary for the younger galaxies.

\section{Summary}
\label{sec:sum}

We have studied the multi-wavelength properties of a sample of $\sim$
200 galaxies with spectroscopically confirmed redshifts at $2.08\leq z\leq 2.51$, 
to study the validity of commonly used dust correction recipes and to compare SFRs inferred from the {\halpha} line luminosity, the UV continuum
luminosity, and {\em Spitzer}/MIPS 24\,$\mu$m measurements that are
converted to the total IR luminosity. In this study, we benefit from using the broad-band photometry excess techniques to determine the {\halpha} fluxes and, hence, conduct a large sample of $z\sim 2$ {\halpha} measurements that is immune to uncertainties in the spectroscopic slit-loss corrections.
The galaxies' properties are determined from stellar population model fitting to the rest-frame UV
through near-infrared data. The main conclusions are as follows.

\begin{itemize}

\item By investigating several recipes for dust correcting the nebular emission
lines, we find that assuming the same color excess of the ionized gas
and the stellar continuum (i.e., $E(B-V)_{\rm neb}=E(B-V)_{\rm stel}$), and
assuming that the Calzetti attenuation curve applies to both, results in the
best agreement between SFR({\halpha}) and SFR(UV).
If we assume the Calzetti curve to both the stellar and nebular regions but use
  $E(B-V)_{nebular}$=2.27~$E(B-V)_{stellar}$, the corrected SFR({\halpha}) measurements
  are inconsistent with the corrected SFR(UV)s at $\sim 5\sigma$ level, averaged on the five bins of SFR(UV) (see the right panel of
    Figure~\ref{fig:sfrhauv}).

\item Using the available {\em Spitzer}/MIPS data for $\sim$ 100 galaxies
  in our sample, we derive an observed L$_{\halpha}$-to-SFR$_{total}$
  conversion factor of $(7.5\pm1.3) \times 10^{-42}$ $\msun$~yr$^{-1}$~erg$^{-1}$~s.
  This calibration is independent of any assumptions on the dust
  correction and can be used to convert the observed (extincted)
  {\halpha} luminosity to a bolometric SFR when 
  no dust attenuation
  measurements for the {\halpha} luminosity is available. 
  
\item By comparing the stacks of SFR(UV)+SFR(IR) with the SED-inferred
  SFRs that are corrected for dust by the locally-derived Calzetti 
  curve we conclude that a steeper attenuation curve (such as an SMC curve) 
  may be necessary for galaxies younger than 100\,Myr, as previous studies \citep{reddy06b,reddy10,wuyts12} have found.
  We find that applying the Calzetti curve to the stacks of all galaxies, including
  the young ones, results in SFR(SED)s that are inconsistent with the SFR(UV)+SFR(IR) at $\sim 2\sigma$ significance (the SFR(SED) in three mass bins of Figure~\ref{fig:sed_bol} are inconsistent with SFR(UV)+SFR(IR) by 1, 2, and 3$\sigma$, from the lowest to highest mass bins respectively).
  The young galaxies have redder UV slope and at the same
  time lower bolometric SFR compared to their older counterparts. As
  a result, applying the Calzetti curve to 
  the young galaxies overestimates the SED-inferred SFR when compared with 
  their bolometric SFR. The Calzetti attenuation curve shows a good overall 
  agreement for galaxies older than 100\,Myr.

\end{itemize}

A more detailed investigation will include direct tracers of nebular
dust extinction (i.e., the Balmer decrement).  In the future studies,
we plan to investigate this aspect with the MOSFIRE Deep Evolution
Field (MOSDEF) survey, which uses the near-IR multi-object
spectrograph MOSFIRE on the Keck~I telescope to obtain spectroscopic
measurements of the nebular emission lines for a sample of $\approx
1500$ galaxies \citep{kriek14}.

IS thanks Valentino Gonz\'{a}lez for useful discussions. Support for IS is
provided through the National Science Foundation Graduate Research
Fellowship under Grant No. DGE-1326120. NAR is supported by an Alfred
P. Sloan Research Fellowship.

\end{document}